\documentclass[fleqn,usenatbib]{mnras}

\usepackage[T1]{fontenc}
\usepackage{ae,aecompl}

\usepackage{newtxtext,newtxmath}

\usepackage{graphicx}	
\usepackage{amsmath}	
\usepackage{tablefootnote}	

\title[Creation of the chemical bimodality]{Fundamental mechanism of the
creation of chemical bimodality in the Milky Way disc in the cold accretion theory}

\author[M. Noguchi]{
Masafumi Noguchi $^{1}$\thanks{E-mail: noguchi@astr.tohoku.ac.jp (MN)}
\\
$^{1}$Astronomical Institute, Tohoku University, 6-3, Aramaki, Aoba-ku, Sendai, Miyagi, 980-8578, Japan
}

\date{Accepted XXX. Received YYY; in original form ZZZ}

\pubyear{2020}

\begin{document}
\label{firstpage}
\pagerange{\pageref{firstpage}--\pageref{lastpage}}
\maketitle

\begin{abstract}
	Chemical bimodality of the Milky Way (MW) disc stars constitutes one of the most
	remarkable properties of MW. The cold accretion theory for the cosmological gas 
	accretion
	provides one viable explanation to this phenomenon.
In this scenario, the rapid cold-mode accretion in the early epoch creates the first generation stars 
	relatively rich in $\alpha$-elements (O,Mg,Si,S,Ca,etc) and
	later cooling flow produces iron-rich second generation stars, creating the bimodality in  the [$\alpha$/Fe] ratio.
	 We employ a cosmologically motivated chemical evolution model for disc galaxies to
	 elucidate the role played by type Ia supernovae (SNIa), which serve as the major source of iron, in the creation of the bimodality.
	 To this end, we divide SNIa into two groups, 
	those formed from the 1st generation stars (the first SNIa)
	and those formed from the 2nd generation stars (the second SNIa).
	The model with the first SNIa suppressed during the {\it second} star formation stage 
	produces stars 
	having high [$\alpha$/Fe] in the early phase of this stage, whereas
	the model which prohibits the second SNIa produces high 
	[$\alpha$/Fe] stars in the late phase. Both models fail to create a well-defined bimodality.
We thus conclude that the cooperation of the first and the second SNIa 
	plays a crucial role in creating the bimodality by maintaining rich iron content 
	in the interstellar gas throughout the second star formation stage.

\end{abstract}




\section{introduction}

Detailed spectroscopic studies in recent years suggest 
the existence of two distinct types of stars with different chemical compositions 
in the Milky Way disc \citep[e.g.][]{ad12,ha13,be14,ha15}.
This bimodality is most clearly represented in the distribution
of long-lifetime stars in the abundance diagram which plots [$\alpha$/Fe] ratio against
[Fe/H]. Although this phenomenon is one of the most notable chemical properties 
characterizing the Milky Way and should give an important clue to the formation process of MW, its origin 
remains a long-standing riddle.
In order to tackle this problem, \citet{ch97} devised a phenomenological model, in which the gas infall 
comprising two distinct episodes is specified.
This approach is also taken in a recent modelling by \citet{sp19}, which 
gives a fair reproduction of the observed abundance pattern. 
Recent studies by \citet{li20a,li20b} also discuss the bimodal abundance distributions and cosntruct chemical
evolution models involving two gas accretion events to explain them.
In these studies, however,
the gas infall history is given by particularly parameterized formular and the best model
is searched by adjusting model parameters. Therefore, the fundamental mechanism that 
underlies the creation of the chemical bimodality remains unclear.

\citet{no18} showed that the cold accretion hypothesis 
 , which is deduced from recent large-scale cosmological simulations, 
 naturally implies a two-stage gas accretion,
 giving one promising solution to this problem.
 These simulations 
 \citep[e.g.][]{fa01,ke05,de06,oc08,va12,ne13}
 trace the thermal history of the cosmic gas as it is incorporated into growing 
 dark matter halos in the cold dark matter universe using hydrodynamical code.
 Specifically, in most of these simulations, the gas component experiences adiabatic heating
 and cooling, shock heating, inverse Compton cooling off the
 microwave background, and radiative cooling via free-free emission,
 collisional ionization and recombination, and collisionally excited
 line cooling. Those simulations provide the most realistic picture of the thermal and 
 dynamical properties of the halo gas in forming galaxies, thereby serving as the best 
 starting point for examining the further evolution of the baryon content in individual 
 galaxies. A particularly important finding of these numerical works is that the cosmic gas
 captured by dark matter halos accretes to the forming galaxies without being heated-up
 by shock waves as envisaged in the classical picture
\citep[e.g.][]{re77},
 but in unheated states 
 when the halos are not massive enough (namely, cold accretion).
 This finding has a large impact on various aspects of galaxy evolution, 
 especially in its early stage \citep{de09}.
 The spirit of \citet{no18} was to apply this cold-accretion hypothesis to MW and to
 interpret chemical properties of individual stars as we observe today in a cosmological perspective.

In the cold-accretion paradigm, the sift of the gas accretion mode
as the halo grows in mass
 leads to the separation of the galactic star formation process into 
two stages \citep{no18}. Initially, the cold-mode gas accretion causes a rapid star formation for 
a couple of Gyrs (the first SF), leaving stars containing copius $\alpha$-elements 
synthesized by type II supernovae (SNII). 
When the accretion process is switched to the cooling flow of the shock-heated 
halo gas (the hot mode) 
\citep[e.g.][]{re77},
the second episode of star formation (the second SF) starts, continuing to the present epoch. 
During this phase, iron-rich stars are
formed. The gap between the free-fall time and the radiative cooling time causes a dormant 
period between two phases, in which  the gas accretion is much suppressed. 
Such a possibility is also suggested by \citet{bi07} in their spherical accretion models 
without star formation.

\citet{no18} succeeded in reproducing the stellar abundance distribution broadly consistent 
with the obervational data for the solar neighbourhood \citep[e.g.][]{ha13} as well as 
the inner and outer discs of MW obtained by the APOGEE project
\citep{ma17}. 
These global features over the wide area of MW are manifested more clealy in the most recent 
APOGEE DR16 data \citep{qu20} showing the bimodality now extending towards the innermost regions,
 lending further credence to the scenario of \citet{no18}.
Moreover, this scenario may also be applicable to other disc galaxies in general.
Indeed, M31, the most massive disc galaxy in the Local group, has a hint of two-stage 
star formation as revealed by color-magnitude diagram analysis of individual stars \citep{wi17}.
Stellar population sysnthesis applied to a number of MW-sized disc galaxies \citep{go17} also suggests 
the existence of two star formation episodes in those galaxies.
These observational studies seem to give a further support to the picture of massive disc galaxy
evolution given in \citet{no18}, suggesting its universality beyond MW.

Although \citet{no18} proposes a promising scenario to explain the observational data,
the key mechanism underlying the bimodality is yet to be clarified.
In \citet{no18}, the high [$\alpha$/Fe] metal-poor stellar population formed in the first SF is
reasonably explained as
the result of relatively rapid star formation which ends before SNIa start to explode massively.
It was observed in the simulation that the interstellar gas
in the second SF epoch remains constantly low in [$\alpha$/Fe] ratio and this fact was used to
explain the formation of a low [$\alpha$/Fe] metal-rich population distinct from the first stellar group,
namely a bimodality. 
Although the low [$\alpha$/Fe] status in the second SF epoch is essential in creating a bimodality, 
how such a state is realized was not clarified. Another unsolved problem is what influence the gas
which remains after the first SF is over has on the gas metallicity in the second SF epoch. 
The remaining gas will mix the fresh gas supplied by the second accretion and may have nonnegligible influence on the gaseous metal content in the second SF epoch.

Type Ia supernovae (SNIa) are believed to be the main supplier of iron in the interstellar gas
\citep[for recent references, e.g.][]{ch97,ha13,ha15}.
In the model of \citet{no18}, SNIa are produced in both the first and the second SF episodes
and both of these SNIa groups can affect the metallicity of the interstellar gas and the stars that formed from it in the second SF epoch.
Because such a grouping of SNIa associated with the separation of star formation into two
distinct epochs is a unique situation only realized in the cold-accretion paradigm, 
it is worthwhile to scrutinize the role played by each SNIa group in the emergence of the chemical bimodality.
To this end, we divide SNIa into two groups according to the birth epoch of their progenitors
and employ a suite of evolution models, in some of which  the effect of either SNIa group is
artificially switched-off. The fiducial model is essentially the same as the model adopted in \citet{no18}.
By intercomparing their behaviour, we can clarify how the two groups cooperate in keeping the high iron abundance
throughout the second stage and thus creating a well-defined bimodality.

	Section 2 gives a brief desciption of the models, while the main results are given in section 3.
Section 4 discusses the limitation of the present work and conclusions are summarized in section 5.

\section{Models}

We use the same evolution code as employed in \citet{no18}.
Briefly, we divide the whole galactic disc into a series  of concentric annuli
and calculate the evolution of each ring under the accretion of gas from the halo.
The cooling rate of the gas which determines the accretion rate is calculated 
assuming the collisional ionization equilibrium. 
The prediction based on the cold accretion theory \citep{de06} is used to
set the epoch dividing the early cold-accretion phase and the late cooling-flow phase.
In the former phase, the halo gas is assumed to accrete to the disc plane with the free-fall time, while the gas accretes with the radiative cooling timescale in the later phase.
The accreting gas is assumed to be pristine (i.e., free of any metals).
It should be noted that, as stated in section 1, the accretion history thus determined is based on the 
results of detailed cosmological simulations for the CDM scenario
 \citep[e.g.][]{fa01,ke05,de06,oc08,va12,ne13}
 , which provide reliable thermal history of the cosmic gas, so that the
time-scales and intensities of the two
 infall episodes, and time gap between them are automatically determined with sufficient physical ground.

We include the star formation process from the accreted gas (i.e., interstellar gas), 
the energy injection and metal enrichment
from SNII and SNIa. 
The former explode instantly 
when their progenitors are formed and return 3.81 ${\rm M}_{\odot}$ of $\alpha$-elements 
and 0.094 ${\rm M}_{\odot}$ of iron.
These are the yields per one SNII averaged over the Salpeter initial mass function.
SNIa explodes with a certain delay which is specified by the delayed time
distribution (DTD) taken from \citet{ma10},

$${\rm DTD}(t)=1 \times 10^{-3} {\rm SN \, Gyr}^{-1} {\rm M}_{\odot}^{-1} t^{-1.1}$$

\noindent{where $t$ denotes the time in Gyr that elapsed since the progenitor formed.}
We assume that DTD=0 for $t<0.3$ Gyr. 
 \citet{ma10} favoured the minimum delay time of 0.1 Gyr based on the galaxy cluster data,
 while \citet{ho15} found that adoption of 0.5 Gyr best reproduces the [$\alpha$/Fe]-[Fe/H]
 diagrams of the local dwarf spheroidal galaxies. The present minimum delay time of 0.3 Gyr
 is the mean of these two values.
Each SNIa returns 0.438 ${\rm M}_{\odot}$ of $\alpha$-elements and 0.63 ${\rm M}_{\odot}$ of iron.
Thus, SNII are the main source for $\alpha$- elements while iron is mostly provided by SNIa.
The $\alpha$-yield for SNII and iron-yield for SNIa used here are respectively increased by 20 percent and 
decreased by 15 percent from the yields given in \citet{fe02}, 
which themselves are based on \citet{th96} for SNII and \citet{th86} for SNIa.
	.
These minor changes are within the estimated uncertainties in the current stellar evolution models \citep[e.g.][]{ro10}, and do not affect
any conclusions of the present study.

In order to mimic MW, we choose the present virial mass of the halo, 
$M_{\rm vir}=1.23\times10^{12}{\rm M}_\odot$.
Setting the supernova feedback efficiency to $\varepsilon=0.15$ leads to 
the disc size and mass nearly consistent with the observed values for MW
{\citep{no18}.
We consider the annulus with $7 {\rm kpc}<r<9 {\rm kpc}$, where
$r$ is the galactocentric distance, as a representative region for the
solar neighbourhood. Fig.1 shows the gas accretion history for this selected region. 
In this MW-analog, the rapid free-fall accretion (i.e., cold-mode accretion) 
continues up to $\sim 8$ Gyr, after which it is taken over by the cooling flow, continuing to
the present epoch.
Because of the gap in the accretion timescale at the transient epoch, a hiatus about 1 Gyr long
appears after the end of cold accretion. 
Detailed mechanism for this two-stage accretion is discussed in \citet{no18}.

We run one fiducial model and four artificial models  with different settings for the included
processes as summarized in table 1. They are classified according to two kinds of options.
First is if the metal enrichment by the first and/or second SNIa is included in the second epoch (i.e. $t$ > 8.6 Gyr)
or not (2nd and 3rd columns). 
The first SNIa here is defined as those
SNIa originating in the stars formed in the first epoch of star
formation (before $t$=8.6 Gyr). The second SNIa refers to 
those SNIa formed from the second 
generation stars created after $t$=8.6 Gyr.
 Second option is if the metal content ($\alpha$ elements and iron) 
in the interstellar gas or the entire gas (including metals) is removed or not at the beginning of the second gas accretion 
(i.e. $t$=8.6Gyr) (3rd and 4th columns). Note that the fiducial model is essentially the same as the
model discussed in \citet{no18}.

\begin{figure}
	\includegraphics[width=1.0\linewidth]{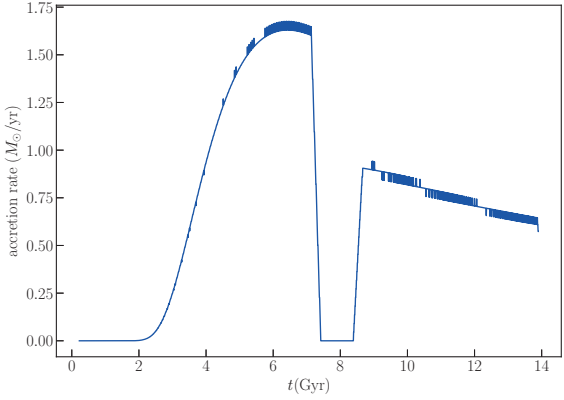}
    \caption{
	Gas accretion rate plotted against the cosmological time,
 $t$. All the models studied in this 
	work have the same 
	accretion history.
	The running mean over 40 successive time steps (0.28 Gyr) is plotted to suppress
	the short timescale fluctuations arising 
	in the model calculation.
	}
    \label{Fig.1}
\end{figure}

\begin{table}
	\caption{Model setting.}
	\label{tab:example}
	\begin{tabular}{lcccc}
		\hline
		Model & SNIa(1st)
		& SNIa(2nd)
		& metal reset
		& gas reset \\
		\hline
		1(fiducial) & Yes & Yes & No & No \\
		2 & No & Yes & No & No \\
		3 & Yes & No & No & No \\
		4 & Yes & Yes & Yes & No \\
		5 & Yes & Yes & Yes & Yes \\
		\hline
	\end{tabular}
\end{table}

\section{results}

Before detailed discussion of the present results, we first show the distribution
of the abundance ratios of the fiducial model (Model 1) in Fig.2 and compare it with
the observation for the solar neighbourhood by the APOGEE project \citep{ma17}.
Incidentally, this model gives the present star formation rate density of 
$3.2 \times 10^1 {\rm M}_\odot/{\rm Gyr}/{{\rm pc}^2}$
for the solar neighbourhood.
The model produces 
two distinct stellar groups, one with high [$\alpha$/Fe] and low [Fe/H]
and another with low [$\alpha$/Fe] and high [Fe/H] in agreement with the APOGEE observation
and other several observational results \citep[e.g.][]{ad12,be14,ha15,ha16}.
The peak locations of respective groups and the total [$\alpha$/Fe] distribution are in reasonable agreement with the observations considering the idealized nature of the present evolution model.
Some discrepancy is seen for more detailed features. The model does not develop
two parallel sequences for [Fe/H]>-0.3 so clearly as reported by \citet{ni20}, with its $\alpha$-rich
branch too weak for this metallicity range.  Recent observations  \citep[e.g.][]{si18}
show overlap also in [$\alpha$/Fe] ratios and stellar ages of the two groups that the 
present model does not produce (the latter by construction).
Some recent studies \citep[e.g.][]{sp19} try to reproduce these detailed features by exploring the parameter space of the chemical evolution models to obtain the best fit values. 
In contrast to these studies, the aim of the present study is to clarify
the most basic mechanism underlying the chemical bimodality
in the framework of the cold accretion theory implied by the CDM cosmology. 
This issue  has not been the subject of intensive studies so far.
In this spirit, this study does not aim to reporoduce every detail of the observed abundance patterns
and we define the bimodality as the existence of two distinct stellar groups, one having 
high [$\alpha$/Fe] and low [Fe/H] and another having low [$\alpha$/Fe] and high [Fe/H]
at the locations in the abundance diagram approximately consistent with the observed ones.
It is noted that the bimodality defined in this way persists in the most recent
APOGEE DR16 observations \citep{qu20}.

As discussed intensely in the literature \citep[e.g.][]{ch97,ha13,ha15}, 
the production of iron by SNIa is 
regarded as 
the essential ingredient bringing about chemical diversity, especially the iron content, in the formed stars.
Because of their delayed response to the star formation events, a rapid star formation 
generally leads to the production of stars with high [$\alpha$/Fe] ratios due to poor
iron content in the parent interstellar gas.
In contrast, a slow star formation with a longer timescale than the delay time produces 
 high [$\alpha$/Fe] stars in early times but starts to create iron-rich stellar populations
 as SNIa explosions become frequent. 

In the fiducial model (Model 1),
the first SF fuelled by the cold accretion produces an 
$\alpha$-element rich stellar population, while the second SF
driven by the cooling-flow produces
iron-rich stars. These processes create the two stellar groups well-separated 
in the [$\alpha$/Fe]-[Fe/H] plane as shown in Figs.2 and 3a.
Delayed explosions of SNIa with respect to SNII
 result into a relatively low ratio of the SNIa rate to the SNII rate
untill the SFR reaches a maximum in the first stage ($t \sim 7.5$ Gyr, Fig.3b), while,
in the second stage, this ratio rises as the SNIa events catch up with SNII explosions  
(the peak ratio at the transition epoch, $t \sim$ 8.6 Gyr, can be neglected because of its short 
duration). This process is reflected in the time variation of elemental abundances 
plotted in Fig.3c. The interstellar gas has a high [$\alpha$/Fe] ratio before $t \sim 8$ Gyr,
whereas this ratio stays at a low level in the second epoch. Because the decline of this 
ratio is abrupt and takes place at the period where SFR is low due to the 
temporal pause of the gas accretion, there appears a gap in the [$\alpha$/Fe] distribution 
between the first and second stellar groups.

\begin{figure}
	\includegraphics[width=1.0\linewidth]{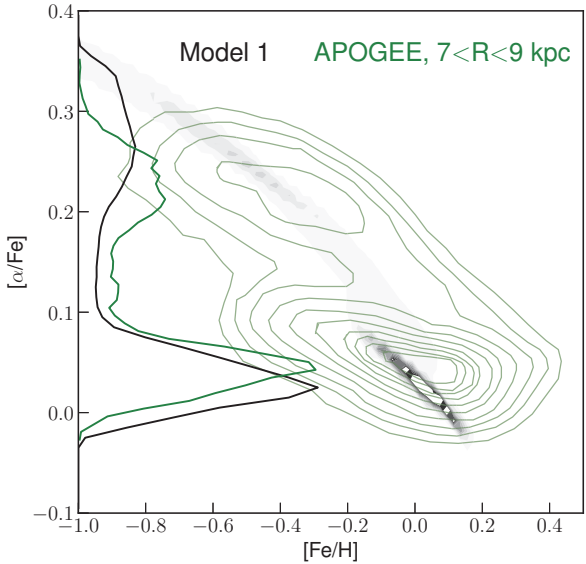}
    \caption{
	Fiducial model compared with the observational data.
	The abundance distribution of the model is indicated by the gray scale image
	whereas green contours show the distribution for the solar neighbour stars 
	obtained by APOGEE \citep{ha16}. The histograms on the left axis show the distribution
	of [$\alpha$/Fe] for the model (black) and the observation (green), normalized so that
	the integrated area is equal.
	\label{Fig.2}}
\end{figure}

\begin{figure*}
	\includegraphics[width=1.0\linewidth]{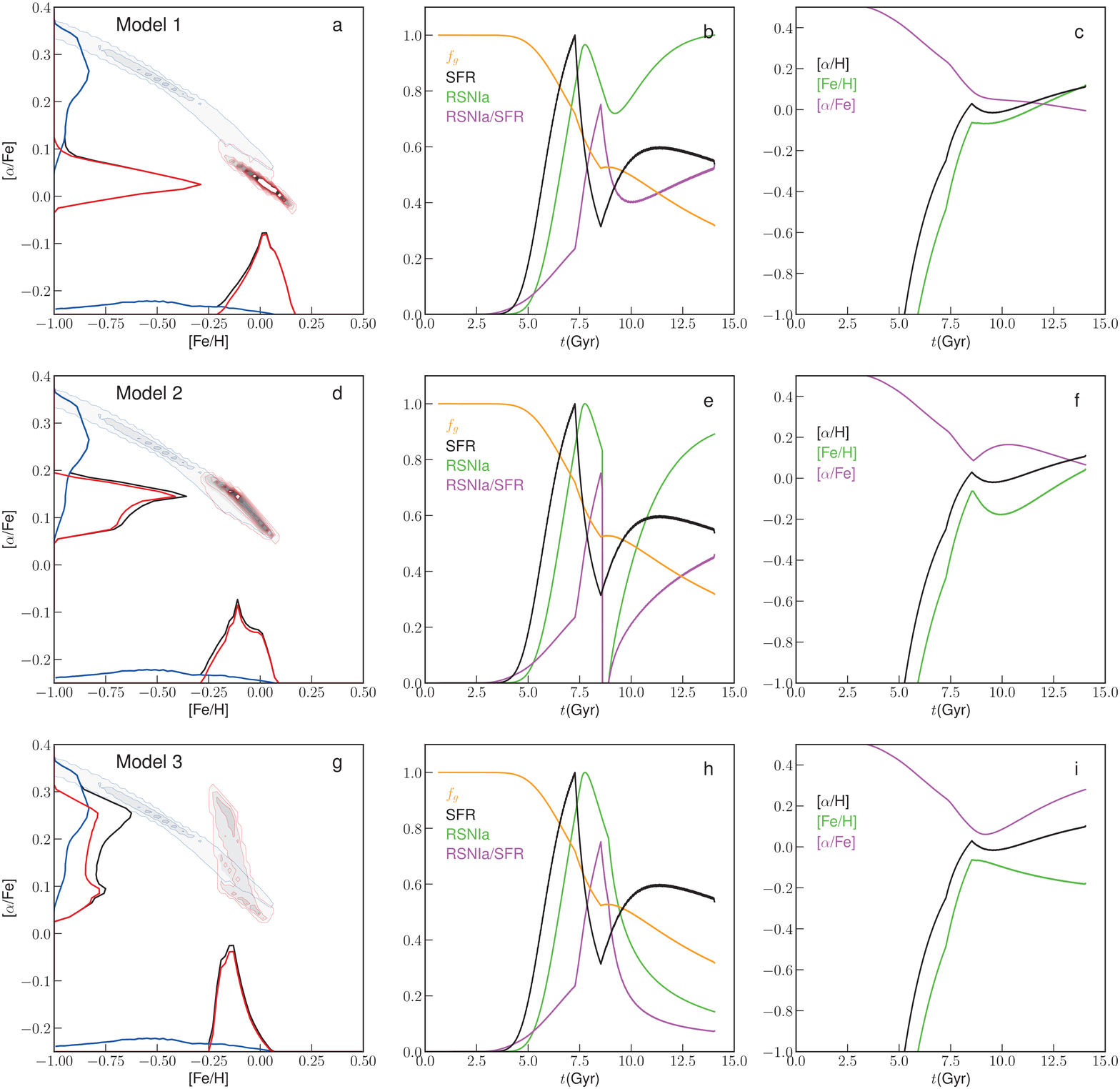}
    \caption{
	Stellar abundance distributions (Panels a,d, and g), the time development
	(Panels b,e, and h), and the chemical enrichment (Panels c,f, and i)
	for Models 1 (upper), 2 (middle), and 3 (bottom).
	In the left panels, the bule and red contours (equally spaced in linear scale)
	 respectively indicate the stellar density
	for the first and second stellar groups divided at $t=8.6$ Gyr, with the total 
	density distribution shown in gray scale. The blue and red histograms 
	on the left axis indicate the [$\alpha$/Fe] distribution for the 
	corresponding groups while the black histogram indicates the total.
	The histograms on the bottom axis indicate the distribution of [Fe/H]
	in the same manner.
	In the centre panels, the gas fraction is calculated as  
	$f_{\rm g}\equiv m_{\rm g}/(m_{\rm g}+m_{\rm s})$ from  
	the interstellar gas mass, $m_{\rm g}$, and the stellar mass, $m_{\rm s}$.
	The rates of the star formation (SFR) and SNIa (RSNIa) are normalized 
	by their maximum values. Ratio of the RSNIa and SFR is given in an arbitrary 
	unit but its normalization is common to the three indicated models to
	facilitate comparison.
         }
    \label{Fig.3}
\end{figure*}

We examine in more detail the role of SNIa in creating this bimodality by comparing the result 
of Model 1 with those of Models 2 and 3, which have different SNIa treatments.

In Model 2, the first SNIa are prohibitted to explode in the second stage (i.e., after $t$=8.6 Gyr).
In this case, the SNIa rate and the SNIa/SNII ratio in the beginning of the second episode are both low, 
being contributed only by the second SNIa (Fig.3e).
Therefore, the [Fe/H] ratio remains smaller than that in the fiducial model (Fig.3f).
Because [$\alpha$/H] stays at a high level from the outset due to quick replenishment
of $\alpha$-elements
from the second SF (Fig.3e), [$\alpha$/Fe] increases during early few Gyrs
in the second stage (Fig.3f). Although the delayed iron enrichment from the second SNIa
lowers [$\alpha$/Fe] in the late phase, the lack of contribution from the first 
SNIa and resulting high [$\alpha$/Fe] ratios during the early phase 
bury the gap between the first and second populations in the abundance diagram
(Fig.3d), leading to an almost continuous sequence. 
Although the histogram in Fig.3d may indicate a bimodality in [$\alpha$/Fe],
the gray scale map  shows that the bimodality in this model is significantly degraded compared 
with that in the fiducial model. This is because the two peaks in 
the [$\alpha$/Fe] distribution are more closely located and the second generation stars  
exhibit a broader distribution. 
Model 2 thus demonstrates that the iron enrichment from the first SNIa extending into the second 
stage is indispensable
in establishing a well-defined bimodality.

Then what about the role of the second SNIa? Model 3 was run with the effect
of the second SNIa totally cutoff, but the first SNIa are allowed to explode
throughout the whole calculated period.
In this case, the stars formed in the early phase of the second epoch have low 
[$\alpha$/Fe] values owing to the delayed action of the first SNIa, leading to their location
 well-separated from the first stellar group as seen in Fig.3g.
In the later phase, however,
RSNIa and hence the SNIa/SNII ratio decrease (Fig.3h) as the contribution from the fisrt SNIa decays
with no subsequent replacement 
from the second SNIa. As a result, the [$\alpha$/Fe] ratio returns back to high values (Fig.3i),
leading to a quite different abundance pattern from the fiducial model (Fig.3g).
This demonstrates that the first SNIa alone cannot produce the bimodality.

In summary, the first SNIa contribute to maintain [$\alpha$/Fe] low 
in the early phase of the second SF, whereas 
the second SNIa keep the low level of [$\alpha$/Fe] in the later
half of the second SF period. The two SNIa groups thus cooperate 
to keep [$\alpha$/Fe] sufficiently low throughout the second SF,
leading to a well-defined bimodality.

\begin{figure*}
	\includegraphics[width=1.0\linewidth]{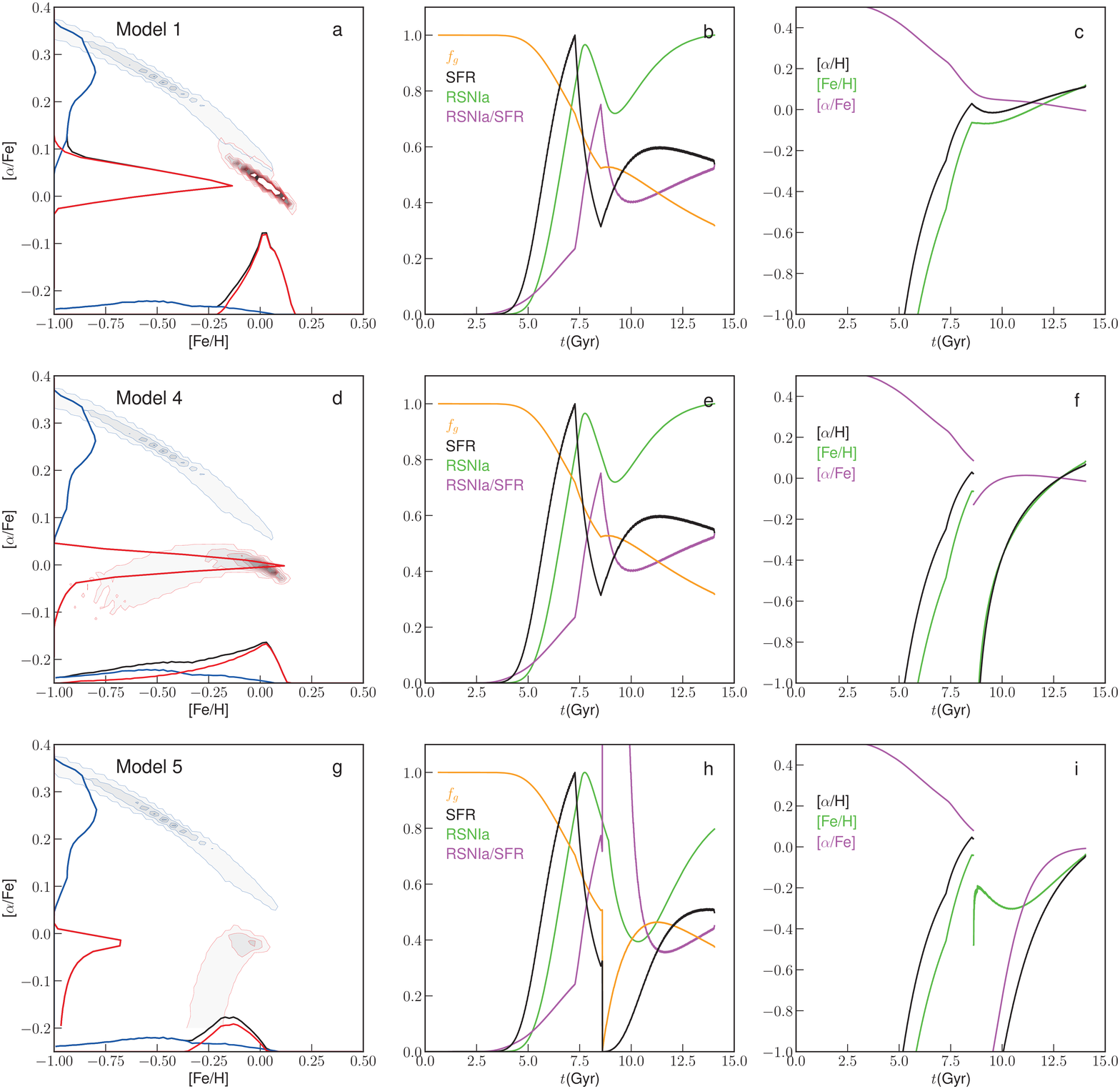}
    \caption{
	    Same as Fig.3 but for Models 1 (upper), 4 (middle), and 5 (bottom).	
	}
    \label{Fig.4}
\end{figure*}

Another interesting question is to what degree the interstellar gas remaining at the dormant period  
contributes to the establishment of the chemical bimodality.
As shown in Fig.3, considerable amount of the interstellar gas (with the gas fraction 
$f_{\rm g} \sim 50$ percent) remains after the first 
SF is over, which contains copius iron and has a low [$\alpha$/Fe] ratio.
If this remaining gas is completely turned into stars, they would occupy a non-negligible portion
of the total second stellar population and affect its abundance ratio.
The low [$\alpha$/Fe] of the second generation stars thus may simply come from the abundance
pattern of this remaining interstellar gas. 

In order to explore this possibility, Model 4 was run
with the abundances of $\alpha$-elements and iron reset to zero at $t$=8.6 Gyr, i.e., 
the beginning of the second SF (Table 1). In other words, the interstellar gas 
is preserved but it is now pristine. 
As shown in Fig.4d, this model produces a well-defined bimodality similar to the one in the 
fiducial model (Fig.4a). In this model, the second SF quickly lifts up the [$\alpha$/H] ratio 
as indicated in Fig.4f. However, the iron production by the first SNIa continuing 
from the first episode promptly rises the interstellar [Fe/H] as well.
The vestige of gas 'purification' applied at $t$=8.6 Gyr is quickly erased.
Therefore, the chemical enrichment history in the second episode of this 
model is essentially the same as that in the fiducial model, and the [$\alpha$/Fe] ratio is low from the early phase of this episode as shown in Fig.4f. 

Model 5 embodies a more extreme situation, in which the whole remaining interstellar gas is removed
at the beginning of the second episode (Fig.4h). This model also produces a clear bimodality
in the abundance diagram (Fig.4g), although details are different from the fiducial model. 
Amount of the second generation stars is reduced because of the gas removal and a long tail 
extends to
very low [$\alpha$/Fe] ratios from the clump of the second generation stars.
Despite these subtle differences, Fig.4i reveals essentially the same enrichment history
of this model as those of Models 1
and 4. The second episode of Model 5 is regarded as just another isolated gas accretion 
and star formation 
event independent upon the outcome of the preceeding first episode 
{\it except} that the first SNIa are
existent from the outset. By its own, such a single star formation event would produce high [$\alpha$/Fe]
stars in the beginning, with this ratio decreasing continuously as the star formation proceeds.
In Model 5, however, the extant first SNIa act to lower [$\alpha$/Fe] ratio of the second 
generation stars from the beginning and no $\alpha$-element rich stars are produced (Fig.4i).
Thus a chemical bimodality is created in  fundamentally the same way as in the fiducial model.
By these experiments, we can conclude that 
the gas remaining after the first SF 
plays at most a secondary role in establishing the chemical bimodality.

\section{discussion}

We briefly discuss the limitation of the current study.
 Here, we did not explore a large domain in the parameter space but considered several
 variations around the fiducial model which is
 plausible for the currently observable regions of the 
 MW disc. 
 Although this approach is adequate for the present purpose, 
 it should be cautioned that the above argument concerning the emergence 
 of the bimodality in the presence of two separated gas accretion episodes may not necessarily apply to 
 other regions of the MW disc or 
 other disc galaxies.
 If, for example, the time-lag between the two star formation episodes is much larger than the 
 decay timescale of SNIa, then
the importance of the first SNIa in the second episode will be much reduced, 
and the entire chemical pattern of the disc stars will be just the superposition of the patterns made by two independent star 
formation episodes. The large overlapping of the two patterns expected in such a case 
will destroy the bimodality.

Is it possible that such a situation takes place in reality? 
In this regard, models with more massive virial masses provide useful information
because the gap duration tends to increase for more massive galaxies. A model with 
$M_{\rm vir}=2.48\times10^{12}{\rm M}_\odot$ was run to examin this point.
This model has a gap duration $\sim 2$ Gyr, about twice that in the fiducail model.
With all other parameters fixed the same as in the fiducial model, the shift in [$\alpha$/Fe]
for the two stellar populations increased from $\sim 0.2$ to $\sim 0.3$, although the
stellar distribution on [$\alpha$/Fe]-[Fe/H] diagram remains qualitatively the same, showing 
a clear bimodality. The present total stellar mass in this model is
$1.05\times10^{11}{\rm M}_\odot$
, nearly at the 
upper limit for the observed disc masses. Therefore, there is little possibility that a too large 
accretion gap results into two overlapping stellar distributions in the abundance diagram.
This more massive model likely represents M31 in the Local Group, for which a longer 
star formation gap than that for MW is actually inferred from the color-magnitude 
diagram dating of individual stars \citep{wi17}. Future observation of
the abundance pattern for this nearby galaxy would be interesting, providing 
a powerful diagnostics for the cold accretion model for galactic chemical evolution.

Another situation is the case in which the second episode of gas accretion dominates the first one.
In this case, the stellar population
formed in the second episode outnumbers the first one, and the whole evolution approaches
to that of a single star formation event. The outcome is an elongated stellar disitribution
in the [$\alpha$/Fe]-[Fe/H] diagram (running from the upper left to the lower right) made by the second generation stars with a negligible 
contribution from the first generation.
This situation actually occurs in the far-out region of the MW disc, for example.
Indeed, the abundance diagram for the outermost region (11kpc < $r$ < 13kpc) shown in \citet{no18}
exhibits such stellar distribution, in agreement with  the observations for the corresponding regions of the MW disc \citep{ha15,qu20}.

Bearing in mind the applicability of this cold-accretion-based scenario to other nearby galaxies, 
a preliminary attempt was made in \citet{no18} to clarify the dependence of 
star formation history on the mass of the host galaxy as partly mentioned above.
Figuring out the necessary and sufficent conditions to ensure a clear bimodality is 
desired also in view of a large variety in the properties 
(the halo virial mass, the stellar and gaseous masses, the star formation history, etc) 
of the Local Group disc
galaxies \citep[e.g.][]{wi09,go10,wi17}.

	Finally, putting aside these theoretical issues regarding the global properties of the 
	bimodality, the cause of several discrepancies
	between the present model and the solar neighbourhood observation in detailed features
	deserves further exploration. 
	For example the present study does not attempt to explain the  [$\alpha$/Fe] poor metal-poor population observed in the solar neighbourbood
 \citep[e.g.][]{ad12,be14}.
This population most likely resulted from 
the migration and mix of stars born at different galactocentric distances
\citep[e.g.][]{sc09}. Although the present code, which calculates the evolution of
individual annular disc regions independently on other regions, does not permit 
the inclusion of the radial migration of disk material, its effect on the resultant chemical abundance pattern 
of the disc stars should be further explored before the chemical evolution of MW disc is fully 
understood.
	Such studies will not only provide additional constraints 
	on the galaxy formation scenarios but also contribute to improve stellar evolution models.

\section{conclusions}

The whole process envisaged for the creation of chemical bimodality in the MW disc stars
is summarized as follows. The change of the gas accretion from the cold mode to the hot mode 
as MW grows in mass 
induces division of the star formation process into two stages. In the first stage, 
the termination of the cold-mode accretion before the massive occurrence of SNIa leaves 
behind a stellar population relatively rich in $\alpha$-elements.
In the second stage of star formation driven by the cooling flow, the interstellar gas is effectively polluted 
by iron ejected by the SNIa formed from the 1st generation stars in its early epoch
whereas the SNIa originating in the 2nd generation stars mainly contribute to iron
enrichment in its late epoch. This ceaseless supply of iron maintains the iron abundance in the interstellar gas at a sufficiently high level
throughout the entire second stage, so that the 2nd generation stars constitute an $\alpha$-poor
 and iron-rich
stellar population well separated from the first generation stars in the [Fe/H]-[$\alpha$/Fe] plane.

\section*{Acknowledgements}

We thank the anonymous referee for useful comments which helped us improve the paper.

\vspace{16pt}

\noindent{Data availability}

\vspace{6pt}
\noindent{The data underlying this article will be shared on reasonable request to the corresponding author.}






\bsp	
\label{lastpage}
\end{document}